# Layered-to-Spinel Phase Transformation in $Li_{0.5}NiO_2$ from First Principles

C. Komurcuoglu[1], A. C. West[1] and A. Urban[1]

[1]*Department of Chemical Engineering and Columbia Electrochemical Energy Center, Columbia University*

E-mail: au2229@columbia.edu

**Abstract**

The phase transition layered $Li_{0.5}NiO_2$ to spinel $Li(NiO_2)_2$ is a potential degradation pathway in $LiNiO_2$-based lithium-ion battery cathodes. We investigated the mechanism of this phase transformation from first principles. Consistent with experimental observations reported in the literature, our results indicate a high energy barrier for the transformation due to high defect-formation energies, a complex charge-transfer mechanism, and electronic frustration. Our results suggest that partially inverse spinel phases are unlikely to form for $Li_{0.5}NiO_2$, a qualitative difference from the chemically similar $Li_{0.5}MnO_2$, in which the transformation occurs at room temperature. We show that Ni and Li atoms migrate concertedly towards their respective spinel sites for the layered-to-spinel transformation to occur. We investigated the charge ordering in layered phases along the $LiNiO_2$-$NiO_2$ composition line, finding a pronounced impact of the symmetry and space group on the layered-to-spinel transition in $Li_{0.5}NiO_2$. Finally, we evaluated the relative stability of different spinel space groups, finding that previously reported experimental observations are consistent with a temperature-averaged structure rather than the 0 Kelvin ground-state structure of $Li(NiO_2)_2$ spinel.

## 1. Introduction

Nickel-rich cathode materials are promising for the next generation of Li-ion batteries since they reduce the reliance on scarce and geographically localized cobalt. Yet, by increasing the Ni content, the degradation of layered cathodes during charge-discharge cycling becomes more prominent.[1] One of the degradation modes in this Ni-rich cathodes is Li/Ni-mixing, in which Ni atoms occupy Li sites,[2–5] causing a reduced Li mobility and first-cycle capacity.[6] This mode is especially pronounced in the limit of $LiNiO_2$, i.e., the most Ni-rich cathode material, and it is, therefore, a useful model system to understand the underlying atomic-scale mechanisms.

For the compositions $Li_{0.5}TMO_2$ (TM = 3d transition metal), the Li/TM-mixed spinel structure is the ground state phase.[7] The phase transformation from the layered to the spinel structure is a degradation process that proceeds via Li/TM mixing and occurs at room temperature in $Li_{0.5}MnO_2$ but is only observed at high temperatures in $Li_{0.5}NiO_2$.[8,9] During the layered-to-spinel phase transformation, TM and Li atoms migrate to their respective antisites, while the oxygen framework remains unchanged.[10–12] Hence, Li/TM mixing is often thought of as related to TM migration in $Li_{0.5}TMOs$ and the layered-to-spinel transition.

Previous first-principles studies by Reed et al. showed that in $Li_{0.5}MnO_2$, the layered-to-spinel transformation occurs via an intermediate phase, the *partially inverse spinel*,[12,13] which has also been experimentally confirmed.[14–16] Recently, Zhang et al. argued that the partially inverse spinel phase cannot be observed in Ni-rich NCA ($LiNi_{0.8}Co_{0.15}Al_{0.05}O_2$) because of its quick decomposition into the spinel structure.[17] In addition, it is experimentally well established that the layered-to-spinel phase transformation in $Li_{0.5}NiO_2$ occurs above 200°C in the bulk of the material.[18–20] The mechanism of the layered-to-spinel transformation in $Li_{0.5}NiO_2$ and its relationship with Ni migration remain unclear.

This study provides a mechanistic insight into the layered-to-spinel transformation from first principles. In the following, we investigate the impact of charge transfer processes and the different $Li_{0.5}NiO_2$ symmetry space groups on the mechanism of the phase transformation. We determine whether the partially inverse spinel phase is an intermediate towards the spinel. And we discuss the room-temperature ground-state structure of spinel $LiNi_2O_4$ and analyze the charge orderings along the phase diagram of layered $Li_xNiO_2$-$NiO_2$.

## 2. Methods & Structure Models

### 2.1. Phase Transformation in $Li_{0.5}NiO_2$

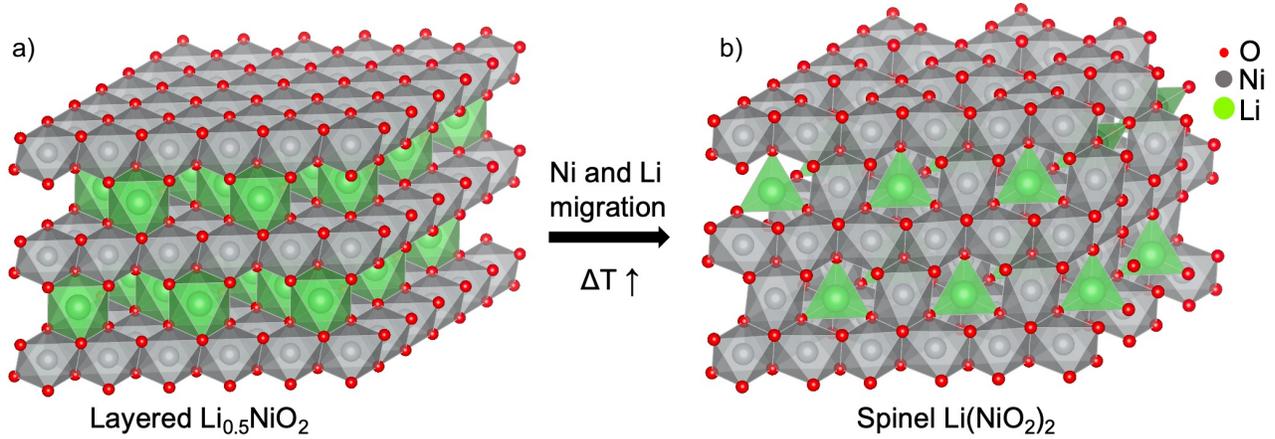

**Figure 1:** Layered and spinel structure models. **a)** In the layered $Li_{0.5}NiO_2$ structure, the Li and Ni atoms are in octahedral sites in separate layers. **b)** In the spinel $Li(NiO_2)_2$ structure, 25% of the Ni atoms have migrated to octahedral sites in the Li layer, and all Li atoms are in tetrahedral sites.

$LiNiO_2$ has a layered α-$NaFeO_2$-type structure. Ideal layered $LiNiO_2$ is comprised of $(NiO_2)_n$ sheets of edge-sharing $NiO_6$ octahedra, in between which Li-atoms are inserted. The $(NiO_2)_n$ sheets are stacked in an ABC sequence, so the structure repeats itself every three layers (O3 type in the notation introduced by Delmas et al.[21]). The Li atoms in the O3 phase can be deintercalated reversibly.[7,21–23] At room temperature, the $LiNiO_2$ structure has been argued to be a thermal average of symmetrically different phases.[24–26] Here, we considered the C2/m, P2c, P2$_1$/c, and R$\bar{3}$m space groups described in more detail in the following section. Layered $Li_{0.5}NiO_2$ (**Figure 1a**) has the same prototype, face-centered cubic (FCC) oxygen framework, and stacking sequence as its fully lithiated $LiNiO_2$ parent.

The $Li(NiO_2)_2$ spinel structure follows the general spinel formula $AB_2X_4$ and possesses the same FCC oxygen framework as the layered $Li_{0.5}NiO_2$. The cation arrangement in the spinel differs from the layered structure wherein, in the spinel, the Li atoms occupy 1/8 of the tetrahedral sites formed by the oxygen framework, and the Ni atoms occupy 1/2 of the octahedral sites.[10,27] The ideal spinel structure has the space group Fd$\bar{3}$m with three-fold rotation axes in the ⟨111⟩ crystallographic directions. Perpendicular to each three-fold axis, layers occupied only by octahedral sites alternate with layers occupied by octahedral sites and tetrahedral sites in a ratio of 1 to 2.[27–29] Although, the experimentally observed space group of $Li(NiO_2)_2$ spinel at room temperature is $Fd\bar{3}m$, Jahn-Teller distortions or other charge orderings might lower the symmetry at low temperatures.[19,27,30] We therefore considered in this study also the space groups Imma, $P4_332$, and $R\bar{3}$. **Figure 1b** depicts a structure model of the ideal spinel structure.

The phase transformation of layered $Li_{0.5}NiO_2$ into spinel $Li(NiO_2)_2$ involves the migration of 1/4 of the Ni atoms in the Ni-layer onto octahedral sites in the Li-layer. Simultaneously, all Li-atoms must migrate from their original octahedral sites into tetrahedral sites within the original Li-layer.

The atomic migration steps required for the layered-to-spinel transition are well established[31] and are shown schematically in **Figure S1**. The first step is the formation of a *Li-Ni dumbbell*.[31] Each octahedral Ni site in the layered structure has one face-sharing tetrahedral site in each of the two neighboring Li layers. A Li-Ni dumbbell forms when the Ni atom migrates into one of the two tetrahedral sites ($Ni_{oct\ Ni-layer} \rightarrow Ni_{tet\ Li-layer}$) while, simultaneously, a Li atom migrates into the second tetrahedral site in the layer opposite the Ni migration direction ($Li_{oct\ Li-layer} \rightarrow Li_{tet\ Li-layer}$). Hence, the Li-Ni dumbbell comprises Ni and Li atoms in tetrahedral sites separated by an octahedral vacancy in the Ni layer ($Ni_{tet\ Li-layer} + Ni_{oct\ vac\ Ni-layer} + Li_{tet\ Li-layer}$). Once the Li-Ni dumbbell has formed, the tetrahedral Ni atom ($Ni_{tet\ Li-layer}$) migrates further within the Li layer onto an octahedral site that is corner sharing with the original octahedral site in the Ni layer from which the Ni migrated ($Ni_{tet\ Li-layer} \rightarrow Ni_{oct\ Ni-layer}$). Simultaneously, a Li atom in the same layer as the migrating Ni atom occupies the now vacant tetrahedral site, forming a Li-Li dumbbell. The spinel motif, i.e., the building block of the spinel structure, is thus composed of a Li-Li dumbbell and a Ni atom on an octahedral site in the former Li layer ($Ni_{oct,\ Li-Layer\ 1} + Li_{tet,\ Li-Layer\ 1} + Li_{tet\ Li-Layer\ 2}$). The spinel structure is obtained when 25% of the Ni atoms have undergone this migration process.

## 2.2. Computational Methods

We used first-principles density-functional theory (DFT) to determine relative phase stabilities and activation energies for migration processes at zero Kelvin. All calculations were performed with the Vienna *ab initio* Simulation Package (VASP)[32–35] and the projector-augmented-wave (PAW) method.[36] We employed the PBE exchange-correlation functional[37] with a rotationally invariant

Hubbard-$U$ correction[38,39] to correct for the self-interaction error of the Ni $d$ states. We considered two different $U$ values previously proposed for Li-Ni oxides, 6 eV[38–40] and 5 eV.[24,41] The value $U = 6$ eV has previously been adjusted to reproduce oxide reaction energies, and we consider it our default. However, since $U = 5$ eV has also been proposed in the literature, we compare trends with both values for sensitive properties. Additionally, we evaluated some of the trends without +$U$ correction (i.e., $U = 0$ eV). A plane wave energy cutoff of 520 eV was used, and Brillouin zone sampling was carried out with a Γ-centered $k$-point mesh with $N_i$ subdivisions along the reciprocal lattice vectors $\mathbf{b}_i$, with $N_i = \lfloor \max(1, R_k|\mathbf{b}_i|) \rfloor$ and $R_k = 25$ Å. The electronic state for selected intermediates was initialized manually. In addition, for all calculations, the geometry of the $NiO_6$ octahedra was initialized with the Ni-O bond length for the expected oxidation state.

To investigate the phase transformation, a 4x4x2 $Li_{0.5}NiO_2$ cell with 32 formula units was utilized (32 Ni atoms) unless otherwise stated. We calculated the formation energy of each symmetrically distinct arrangement of defects (dumbbell or spinel motifs) by systematic enumeration. This allowed us to investigate the stepwise formation of the partially inverse spinel and the spinel starting from the layered structure. The energy of each step in the transformation is

$$E = E_{\text{Defect(x\%)}} - E_{\text{Layered}}$$

where $E_{\text{Layered}}$ is the DFT energy of the ideal layered structure and $E_{\text{Defect(x\%)}}$ represents the energy of a structure with a defect concentration of x%.

Phase diagrams along the $LiNiO_2$-$NiO_2$ composition line were determined by constructing the lower convex hull (LCH) of the formation energies with intermediate compositions. We used a cell with eight primitives $LiNiO_2$ unit cells (2x2x2) for each space group ($C2/m$, $P2_1/c$, and $P2/c$) and scanned for the most stable Li-arrangement by enumerating Li-orderings for each possible Li-concentration, resulting in a total of 156 symmetrically distinct configurations. Crystal structures for additional previously reported Li orderings with larger unit cells were constructed manually. Selected phases with formation energies on or within 7 meV/atom from the LCH were recalculated with larger cells (32 formula units) to obtain converged energies.

*Enumlib* was utilized for structure enumeration.[42–44] Structure analysis and manipulation were carried out with the *Atomic Simulation Environment* (ASE)[45] and the *Python Materials Genomics* (PyMatgen) package.[46]

To calculate oxidation states, we analyzed the atomic magnetization by integrating the difference in electron density of spin up $n_\uparrow(r)$ and spin down $n_\downarrow(r)$ within spheres of increasing radius. The resulting magnetization was used to assign an oxidation state based on the Aufbau principle and ligand-field splitting.

## 3. Results

In the following sections, we first determine the likely Li/vacancy and charge ordering in layered $Li_{0.5}NiO_2$ by revisiting the delithiation of $LiNiO_2$. The most plausible $Li_{0.5}NiO_2$ structures are then taken as the initial structures for modeling the layered-to-spinel transition.

### 3.1. Charge Orderings in the $LiNiO_2$-$NiO_2$ 0K Phase-Diagram & $LiNiO_2$ Space Groups

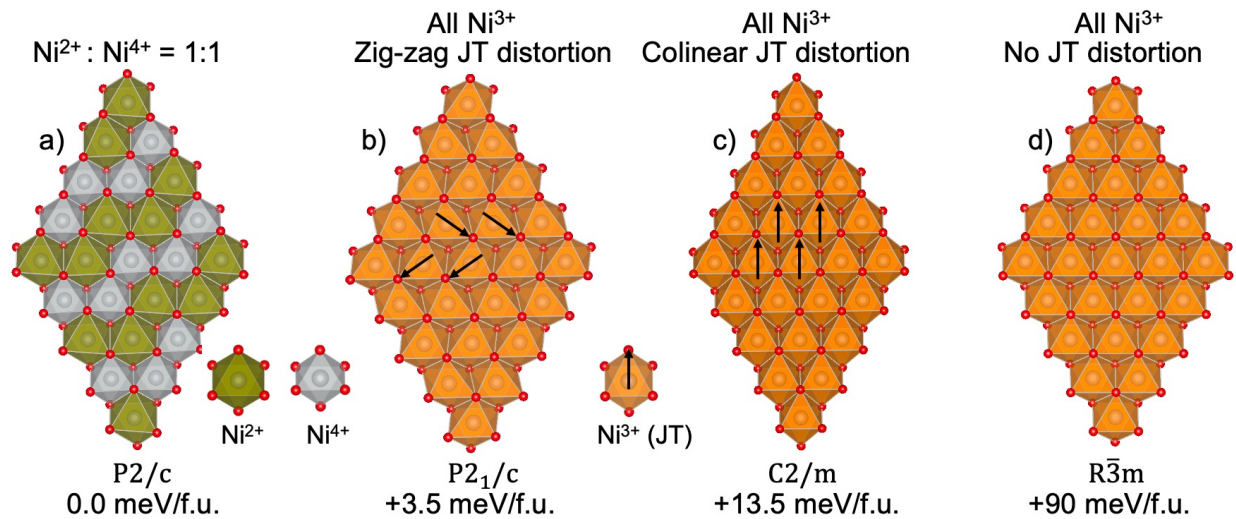

**Figure 2:** Structure models, i.e., the $(NiO_2)_n$ sheets, for four different space groups of layered $LiNiO_2$ and energies per formula unit (f.u.) obtained with PBE+U (U = 6 eV). $(NiO_2)_n$ sheets are shown to visualize the charge ordering and Jahn–Teller (JT) distortions. Li atoms are omitted for clarity. **a)** $P2/c$ structure characterized by charge disproportionation into 50% $Ni^{2+}$ and 50% $Ni^{4+}$. **b)** $P2_1/c$ structure displaying a zig-zag ordering of JT distortions (arrows) with 100% of $Ni^{3+}$. **c)** $C2/m$ structure displaying a collinear ordering of JT distortions with 100% of $Ni^{3+}$. **d)** $R\bar{3}m$ structure with undistorted $Ni^{3+}$ sites.

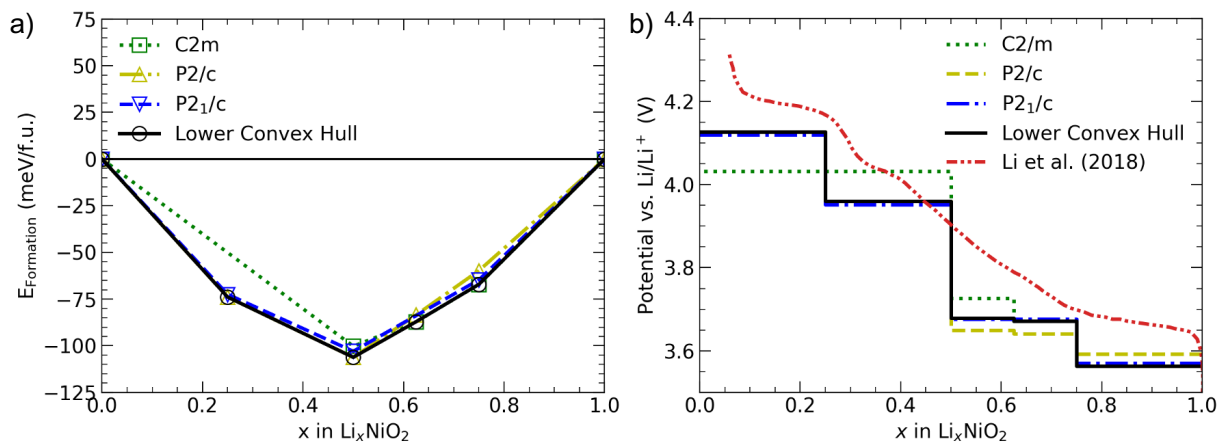

**Figure 3: a)** Zero-Kelvin formation energy of phases along the LiNiO$_2$–NiO$_2$ composition line as a function of the Li concentration for delithiation of initial LiNiO$_2$ structures in the space groups $C2/m$, $P2/c$, and $P2_1/c$. For each space group, only the lower convex hull (LCH) of the formation energies is shown, and the thick black line indicates the overall LCH. The space groups of the partially delithiated Li$_x$NiO$_2$ are shown in subsequent figures. See **Figure S2** for a plot showing the energies of all enumerated configurations. **b)** Voltage profile corresponding to the phase-group-specific and overall LCHs from panel a). A measured room-temperature voltage profile from Li et al.[47] is shown as dashed-dotted line.

**Figure 2** shows the geometry and charge ordering in Ni slabs of fully lithiated LiNiO$_2$ prepared in three different space groups, along with the corresponding relative formation energies predicted by PBE+U (U = 6 eV) calculations. At zero Kelvin, the most stable space group was found to be $P2/c$, followed by $P2_1/c$, $C2/m$, and $R\bar{3}m$. Chen et al.[41] and Radin et al.[24] showed that the order of stability depends on the choice of the +U correction and the functional, and a LiNiO$_2$ stability analysis for different U values is shown in **Figure S3**, which is in agreement with the results by Chen et al.[41]

LiNiO$_2$ with the $P2/c$ space group (**Figure 2a**) is characterized by a charge disproportionation, in which 50% of the Ni atoms are in a 2+ oxidation state and 50% in a 4+ oxidation state. Both Ni$^{2+}$ and Ni$^{4+}$ sites are ideal octahedra, but the Ni$^{2+}$-O distance is ~2.05 Å, whereas the Ni$^{4+}$-O distance is ~1.89 Å.

In LiNiO$_2$ with $P2_1/c$ space group (**Figure 2b**), all Ni atoms are in a 3+ oxidation state. The Ni$^{3+}$ sites are Jahn–Teller (JT) distorted with four short (~1.89 Å) Ni-O bond lengths and two long (~2.10 Å) Ni-O bond lengths. Within the Ni layers, the JT distortions assume a zig-zag ordering.

The C2/m space group (**Figure 2c**) is characterized by a collinear ordering of the JT-distorted $Ni^{3+}O_6$-octahedra, i.e., all octahedral elongations point in the same direction. The Ni-O bond lengths are the same as in the $P2_1/c$ structure.

In the $R\bar{3}m$ structure (**Figure 2d**), all $Ni^{3+}$ sites are ideal octahedra with a Ni-O distance of ~1.97 Å. Although, $R\bar{3}m$ is the experimentally observed space group of $LiNiO_2$ at room temperature, it has previously been argued that the measured structure is a thermal average of structures with lower symmetry.[25,26] This is in agreement with the high relative formation energy predicted by DFT. The $R\bar{3}m$ phase was only included here for the sake of completeness and will not be considered in the following.

**Figure 3** shows the lower convex hull of formation energies for $Li_xNiO_2$ structures obtained via the delithiation of $LiNiO_2$ structures with the three relevant space groups (P2/c, $P2_1/c$, and C2/m). Also shown in the figure are the corresponding zero-Kelvin voltage profiles. The energy differences of the partially delithiated $Li_xNiO_2$ phases with different space groups are, for every Li content, lower than the thermal energy at room temperature (25 meV per degree of freedom).

Voltage profiles for U = 5 eV and uncorrected PBE (U = 0 eV) are shown in **Figure S4**. We find that a +U correction of 6 eV yields the closest agreement with the experimental voltage profile across the entire Li concentration range. Nonetheless, for U = 6 eV, there persists a discrepancy between experiments and computations due to thermal effects.

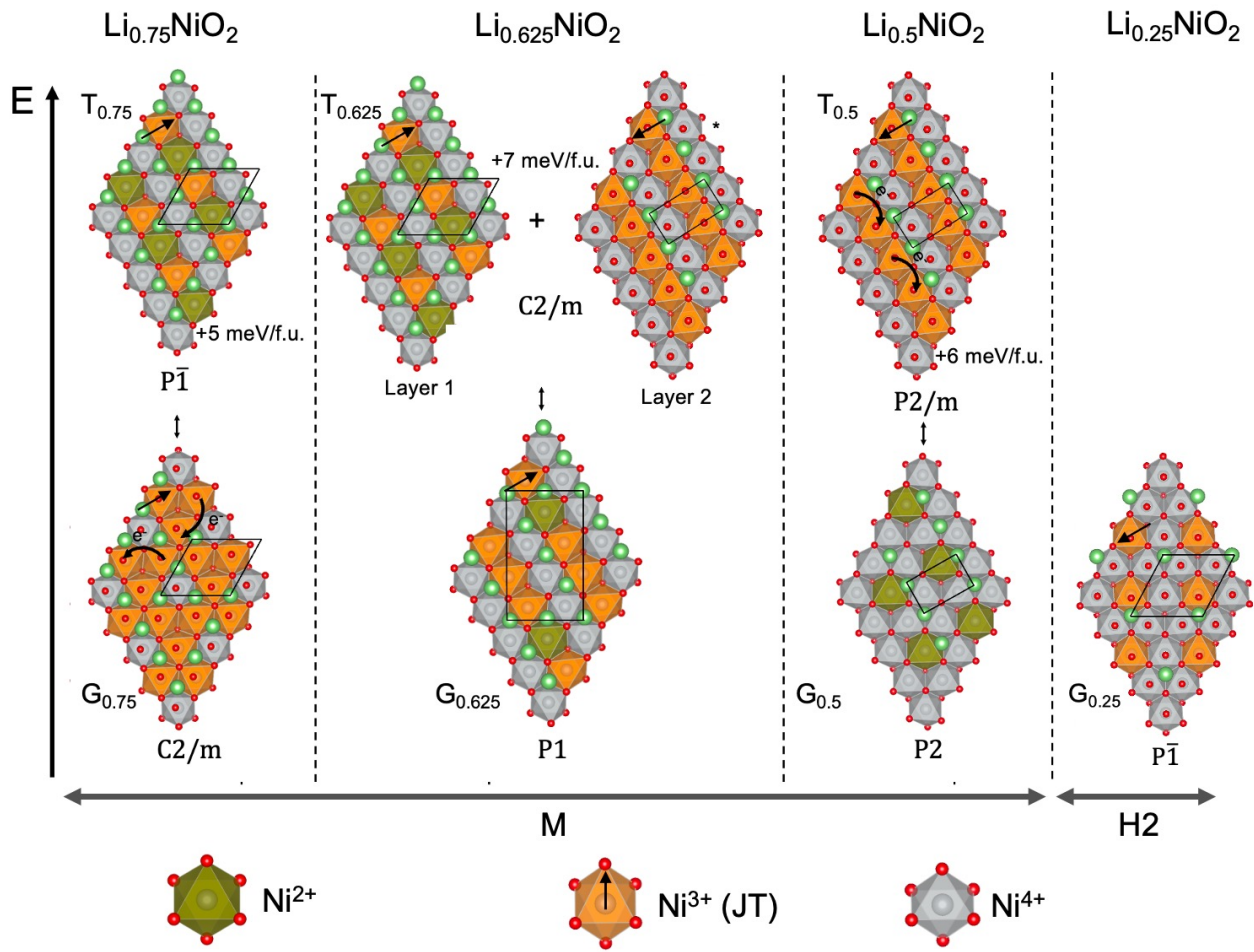

**Figure 4:** Charge and Li/vacancy orderings in $Li_xNiO_2$ on the overall LCH of **Figure 3a** and charge orderings energetically slightly above (<10 meV/f.u.) the LCH. Primitive unit cells are indicated with thin black lines. Curved arrows in the $C2/m$ $Li_{0.75}NiO_2$ structure indicate the electron transfer needed to create the corresponding $P\bar{1}$ ordering. The composition domains of the Monoclinic M, and Hexagonal H2 are shown for illustrative purposes. Basal unit cells are shown with solid black lines. For details about the onset of these domains, see Li et al.[47]

**Figure 4** shows Li/vacancy and charge orderings in $Li_xNiO_2$ along the $LiNiO_2$–$NiO_2$ composition line. Our enumeration identified four stable $Li_xNiO_2$ phases at $x$ = 0.75, 0.625, 0.5, and 0.25 with the same Li/vacancy orderings as previously reported at 0K by Mock et al[48] in their computational study where they used SCAN + rVV10 as the choice of functional. Mock et al. also found a stable ordering at $x$ = 0.4, but our PBE results only predict this ordering to be stable without a +U correction, in agreement with Arroyo y de Dompablo et al.[49] Calculated a/b ratios of ordered stable phases are shown in **Figure S5**

*Charge orderings of the zero-Kelvin phases*

Compositions $Li_xNiO_2$ with formation energies on the lower convex hull are predicted to be thermodynamically stable at zero Kelvin. At $x = 0.75$, we found the most stable charge decoration to be a collinear ordering of JT-distorted $Ni^{3+}$ sites in which all elongations of the $NiO_6$-Octahedra are oriented to a common spatial direction. We refer to this structure as $G_{0.75}$ (ground-state at $x = 0.75$) in the following. The space group of this phase is C2/m, and it contains $Ni^{3+}$ and $Ni^{4+}$ ions in a 3:1 ratio. Computationally, this concentration marks the beginning of the monoclinic domain.[48,50] For a detailed discussion of the experimentally observed onset of the domains, see Li et al.[47] For the same composition, a second configuration with space group $P\bar{1}$, in which 2/3 of the $Ni^{3+}$ atoms have undergone disproportionation into $Ni^{2+}$ and $Ni^{4+}$ ($Ni^{2+}:Ni^{3+}:Ni^{4+}$ ratio 1:1:2), is only 5 meV/f.u. above the LCH and could be thermodynamically accessible at room temperature. We denote this structure as $T_{0.75}$ (thermally accessible at $x = 0.75$).

At $x = 0.625$, we find disproportionation to be preferred at zero Kelvin with the $Ni^{2+}:Ni^{3+}:Ni^{4+}$ ratio 1:3:4 (structure $G_{0.625}$). As discussed by Mock et al.[48] and Peres et al.,[51] the Li ordering at this concentration, can't be described by a single Li-layer, but requires doubling along the axial direction. Hence, the authors identified the structure at this concentration as an O6 structure.[48,52] Also at this composition, we found a thermally accessible structure that is 7 meV/f.u. above the LCH. This structure, $T_{0.625}$, has the space group C2/m and possesses two inequivalent Ni layers: one Ni layer exhibits the disproportionated Li ordering of structure $T_{0.75}$, and in the second Ni layer $Ni^{3+}$ and $Ni^{4+}$ atoms as well as Li and vacancies are ordered in alternating stripes with ratio 1:1, corresponding to a Li content of $x = 0.5$.

A low-energy structure, $T_{0.5}$ (6 meV/f.u. above the LCH), with the overall Li content $x = 0.5$ is obtained when all Ni layers exhibit the $Ni^{3+}/Ni^{4+}$ striped ordering found in every other layer of the $T_{0.75}$ structure. The space group of the $T_{0.75}$ structure is P2/m, and the JT-distorted $Ni^{3+}$ are oriented collinearly, similar to fully lithiated $LiNiO_2$ in the C2/m space group (**Figure 2c**).

Peres et al. first proposed Li orderings at the two Li concentrations $x = 0.75$ and 0.625 based on electron diffraction.[51] Structure $T_{0.625}$ contains two coexisting different Li and charge orderings corresponding to the orderings in $T_{0.75}$ and $T_{0.5}$. Peres et al.[51] by means of electron diffraction data argued that the vacancy orderings in either $T_{0.75}$ and $T_{0.5}$ can be formed in the monoclinic range. XRD and Monte Carlo simulations by Mock et al.,[48] reported the structure with x = 0.625 to exhibit an O6 stacking sequence in which two Li-layers are needed to describe the supercell.

Structure $T_{0.5}$, i.e., a phase with alternating rows of Li atoms and vacancies at $x = 0.5$, is also well established. Experimentally, the space group at this concentration has been found to be P2/m based on Electron diffraction, XRD and Monte Carlo simulations.[48,51] Peres et al. utilizing electron diffraction also obtained the P2/m space group for $T_{0.5}$ with the same Li ordering as that reported here.

However, at zero Kelvin, the DFT ground-state structure at $x = 0.5$ ($G_{0.5}$) is obtained when the $Ni^{3+}$ ions in the $T_{0.5}$ structure disproportionate into $Ni^{2+}$ and $Ni^{4+}$, so that the overall $Ni^{2+}:Ni^{4+}$ ratio is 1:3. This is a delithiation product of the P2/c cell, which is likewise the zero-Kelvin phase of $LiNiO_2$ and contains $Ni^{2+}$ and $Ni^{4+}$ in a 1:1 ratio. The space group of the $G_{0.5}$ structure is P2. We note that the energetic order of the P2 ($G_{0.5}$) and P2/m ($T_{0.5}$) structures is independent of the +U value used with PBE. The disproportionated P2 structure is predicted to be more stable at this concentration at zero Kelvin.

In the following, we will investigate the layered-to-spinel transition initiating from both the P2 and the P2/m structure of $Li_{0.5}NiO_2$. Although the P2/m structure is likely the structure of $Li_{0.5}NiO_2$ at room temperature, based on its agreement with experimental observations, the P2 structure might be stable at low temperatures.

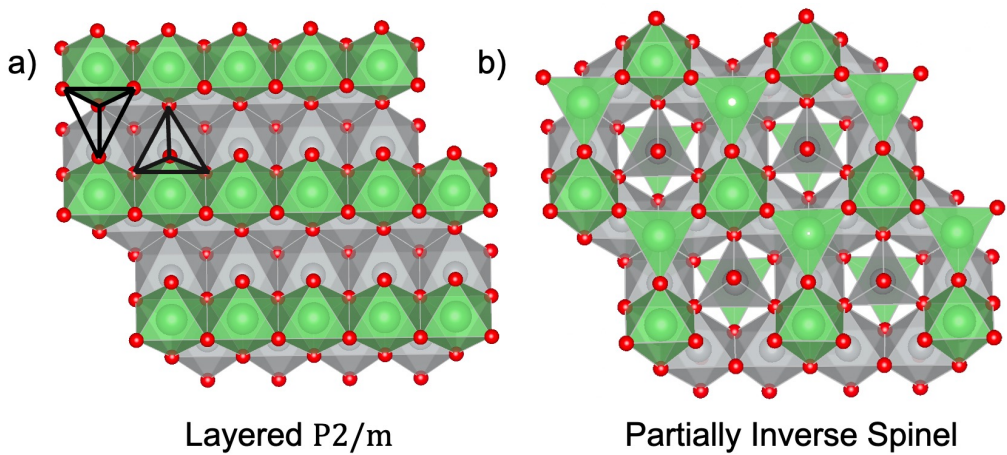

**Figure 5**: Structure models of $Li_{0.5}NiO_2$. **a)** The layered $P2/m$ structure is the room-temperature structure obtained by delithiation of $LiNiO_2$. The tetrahedral sites wherein Li and Ni migrate are indicated with thin black lines. **b)** Partially inverse spinel with 25% of total Ni in tetrahedral sites, 50% Li in tetrahedral, and 50% in octahedral sites.

### 3.2. Layered to Partially Inverse Spinel Transformation in $Li_{0.5}NiO_2$

As discussed in the introduction section, the partially inverse spinel structure is observed as an intermediate of the layered-to-spinel transformation in some lithium transition-metal oxides. In this section, we investigate whether the formation of the partially inverse spinel is plausible for $Li_{0.5}NiO_2$.

**Figure 5 a** and **b** show the P2/m phase of layered $Li_{0.5}NiO_2$ and the partially inverse spinel phase side by side. As discussed in the previous section, the Li ordering in the P2/m layered structure consists of Li-rows alternating with rows of vacancies, and all Ni atoms are in octahedral sites in segregated layers. In the partially inverse spinel, 25% of the Ni atoms are in tetrahedral sites in the Li-layer, and 50% of the Li atoms are also in tetrahedral sites forming Li-Ni dumbbells (see **section 2**) [14,31] The remaining 50% of the Li atoms are in octahedral sites within the Li-layer.

**Figure 6** shows the energetics of the layered-to-partially inverse spinel transformation as a function of the percentage of Ni and Li atoms in the Li-Ni dumbbell configuration. i.e., as a function of the percentage of the layered crystal that has transformed into the partially inverse spinel phase. The x-axis can be understood as an inversion parameter $\lambda$ that indicates the degree to which the layered structure has transformed into the partially inverse spinel.[53]

In the partially inverse spinel, 25% of Ni atoms are in tetrahedral sites in the Li-layer, corresponding to 8 Ni atoms in our $Li_{16}Ni_{32}O_{64}$ (4x4x2). The percentages in **Figure 6** are relative to the eight migrating Ni atoms, and the relative energy was calculated as

$$E_{rel} = \frac{E_{x\%} - E_{layered}}{32 \, Li_{0.5}NiO_2} ,$$

where $E_{x\%}$ is the energy of the layered phase that has undergone x% of the transformation to the partially inverse spinel phase, and $E_{layered}$ is the energy of the defect-free layered structure.

As seen in the figure, the layered-to-partially inverse spinel transformation is an energetically uphill process that requires 245 meV/f.u. (relative to the P2/m $Li_{0.5}NiO_2$ layered structure). Already the formation of one single Li-Ni dumbbell is predicted to be thermodynamically unfavorable and dependent on the oxidation state of the migrating Ni atom, with energetic order ($Ni^{2+}$ in P2) < ($Ni^{3+}$ in P2/m) < ($Ni^{4+}$ in P2/m) and the defect formation energies 0.67, 0.69 and 1.49 eV/defect, respectively, as seen in **Figure S6**. These energies were obtained by calculating the formation energies of a single Li/Ni-dumbbell in the respective $Li_{0.5}NiO_2$ layered structure. The different energies show that charge transfer processes play an important role in the energetics of Ni migration.

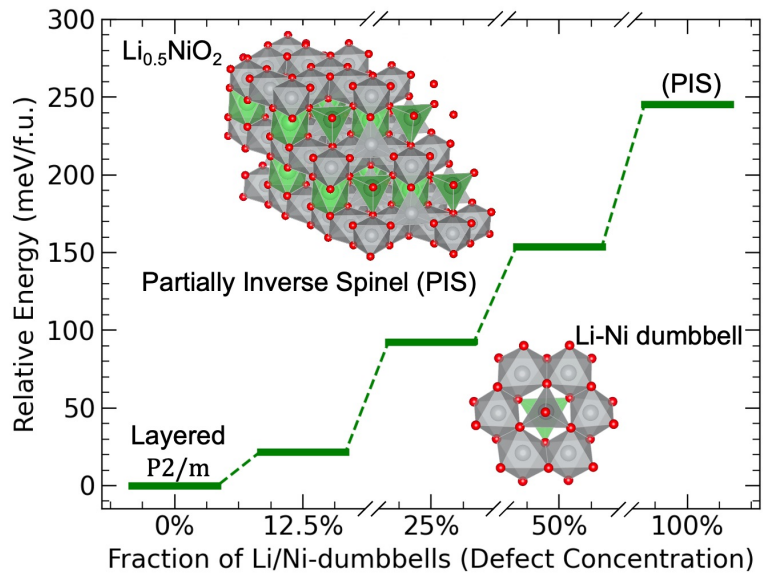

**Figure 6:** Energetics of the layered to partially inverse spinel phase transformation in $Li_{0.5}NiO_2$ as a function of the fraction of Ni atoms in tetrahedral sites (Li-Ni dumbbells). The Li-Ni dumbbell is the building block of the partially inverse spinel. Layered $Li_{0.5}NiO_2$ with the $P2/m$ phase group was the initial structure.

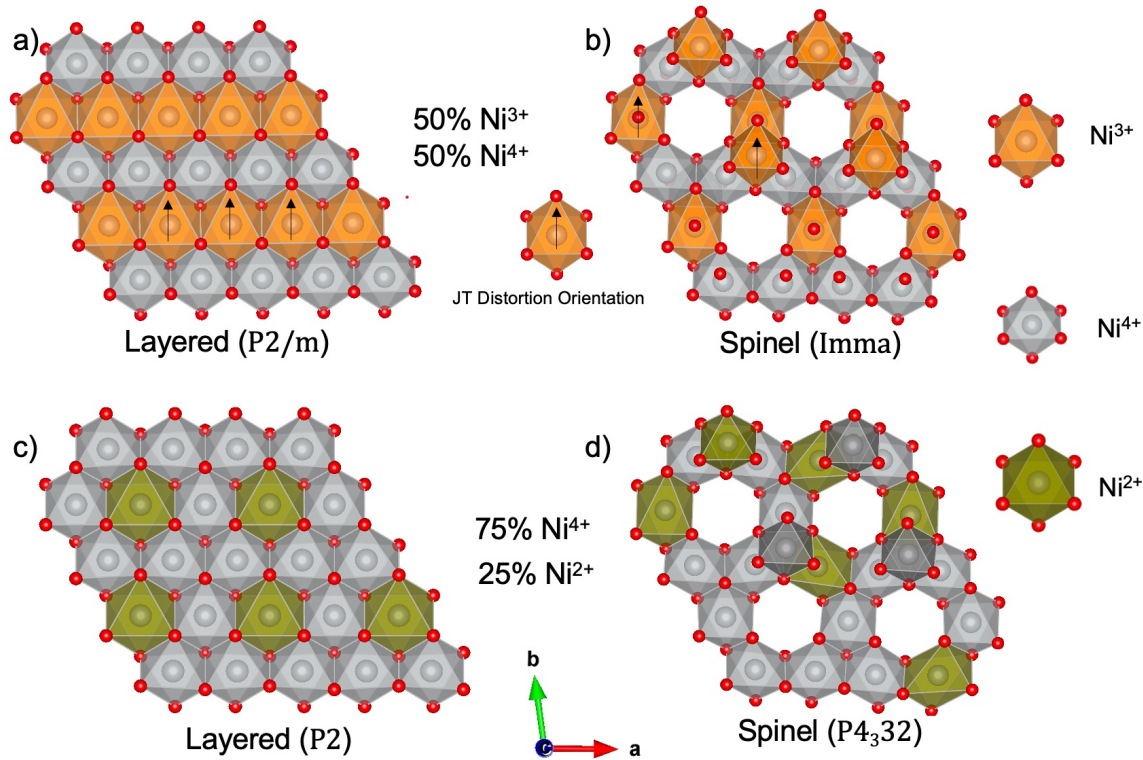

**Figure 7:** Charge and JT-distortion ordering in layered $Li_{0.5}NiO_2$ and spinel $Li(NiO_2)_2$. **a)** In the layered $P2/m$ structure, the JT-distortions of the $Ni^{3+}$ sites are collinear and the Ni atoms are in 3+ and 4+ oxidation states in a 1:1 ratio. **b)** In the related spinel $Imma$ structure, the charge ordering and the orientation of the JT-distortions are maintained. **c)** In the layered $P2$ structure and **d)** the spinel $P4_332$ structure, no JT-distortions are present, and Ni atoms are in 2+ and 4+ oxidation states in a 1:3 ratio. Li atoms are omitted from the figures for clarity.

### 3.3. Layered-to-Spinel Phase Transformation in $Li_{0.5}NiO_2$

The formation of the partially inverse spinel structure is thermodynamically unfavorable and it is thus not an important intermediate of the layered-to-spinel transformation in $Li_{0.5}NiO_2$. Since the spinel structure is overall thermodynamically more stable than the layered structure at the composition $Li_{0.5}NiO_2$, we determined the overall energetics of the spinel formation. We considered both the room-temperature P2/m structure of layered $Li_{0.5}NiO_2$ and the P2 zero-Kelvin ground state as the initial structure for the transformation, as the energetics for the transformation depend on the space group of the layered $Li_{0.5}NiO_2$ structure.

The charge orderings of the P2 and P2/m layered $Li_{0.5}NiO_2$ structures are shown in **Figure 7** together with the charge ordering of two spinel space groups, Imma and $P4_332$. As seen in the figure, the charge ordering and JT distortions connect the layered P2/m and the spinel Imma structures as well as the layered P2 and the spinel P4$_3$32 structures. Like the layered P2/m, in the Imma spinel, the charge ordering consists of 50% JT-distorted Ni$^{3+}$ and 50% Ni$^{4+}$. Ni atoms that migrated from the Ni-layer to the Li-layer are in a 3+ oxidation state. In the P4$_3$32 spinel, the charge ordering consists of 25% Ni$^{2+}$ and 75% Ni$^{4+}$, as is the case for the layered P2.

As described in the methods section and visualized in **Figure S1**, the spinel defect (Ni$_{\text{oct Li-Layer 1}}$ + Li$_{\text{tet Li-Layer 1}}$ + Li$_{\text{tet Li-Layer 2}}$) is the building block of the spinel structure. **Figure 8** depicts the energetics of the layered-to-spinel transformation as a function of increasing spinel defect concentration for the two layered structures. The x-axis indicates the degree to which the layered structure has transformed into the spinel structure. In the spinel structure, 25% of the Ni atoms are in octahedral sites in the original Li-layer, corresponding to 8 Ni atoms in our Li$_{16}$Ni$_{32}$O$_{64}$ structure model (4x4x2). The percentages shown in **Figure 8** are relative to the 8 atoms that migrate, and the relative energy per formula unit was calculated equivalently to the partially inverse spinel structure above

$$E_{\text{rel}} = \frac{(E_{\text{x\%}} - E_{\text{layered}})}{32 \text{ Li}_{0.5}\text{NiO}_2} .$$

As anticipated from the analysis of the charge ordering above, the transformation of the layered P2/m yielded the spinel Imma structure, and the transformation of the layered P2 yielded the spinel P4$_3$32 structure. As seen in **Figure 8**, the transformation of the layered P2/m into the spinel Imma structure requires a lower activation energy (~41 meV/f.u.) than the transformation of the P2 structure (~95 meV/f.u.). The formation energy of the spinel Imma relative to the layered Li$_{0.5}$NiO$_2$ P2/m is -110 meV/f.u. and the formation energy of the spinel P4$_3$32 relative to the layered P2 is -122.57 meV/f.u. which is in agreement with the results by Kuwabara et al.[30]

The transformation of the layered to the spinel structure for different Hubbard-U corrections (+U = 5, 0 eV) is shown in **Figure S7**, and our results show robustness with respect to the choice of the U value with PBE.

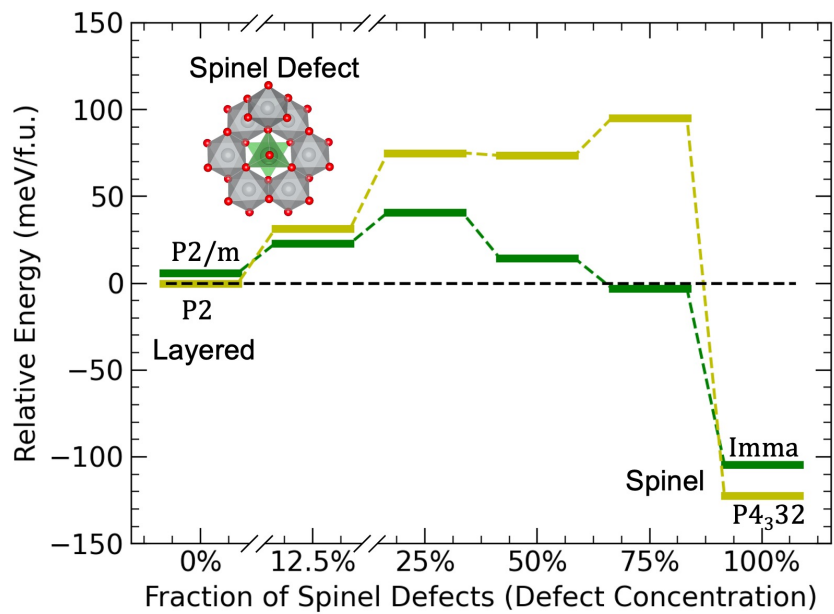

**Figure 8:** Energetics of the layered Li$_{0.5}$NiO$_2$ to spinel Li(NiO$_2$)$_2$ phase transformation for two different initial space groups of the layered structure ($P2$ and $P2/m$) as a function of the spinel defect concentration. The two layered space groups also transform to spinel structures with different space groups ($P4_332$ and $Imma$).

In our analysis of other spinel space groups, we found the most stable space group to be P4$_3$32 followed by Imma (+18.14 meV/f.u.), Fd$\bar{3}$m (+143.32 meV/f.u.) and R$\bar{3}$ (+197 meV/f.u.). In the Fd$\bar{3}$m space group, the ideal spinel, all Ni atoms have an average oxidation state of (3.5+) and all Ni-O bond lengths are identical (~1.91 Å). We found that regardless of the U value in our PBE+U calculations, the P4$_3$32 is the most stable, followed by the Imma structure. Our relative stabilities of the space groups in agreement with those found by Kuwabara et al for PBE + 6 eV.[30] Spinel formation energies for other choices of U values are shown in **Figure S3**.

## 4. Discussion

Using first principles calculations, we investigated the charge ordering in the phase diagram of LiNiO$_2$-NiO$_2$ and the mechanism of the layered-to-spinel transition, considering also the formation of partially inverse spinel. We investigated the relative stability of the spinel structure with different space groups.

We found that the layered-to-spinel transformation is a kinetically hindered and concerted process in which large domains of the layered crystal transform into the spinel simultaneously. The partially inverse spinel is not an intermediate state during the layered-to-spinel transformation, as it requires high energy to form, explaining why the partially inverse spinel is not observed in Ni-rich cathode materials (in contrast to other layered TMOs, such as $Li_{0.5}MnO_2$).

To identify the most likely structure of layered $Li_{0.5}NiO_2$, we modeled the delithiation of $LiNiO_2$ with different space groups. In diffraction experiments, the structure of $LiNiO_2$ appears to have the R$\bar{3}$m space group, with 3+ Ni atoms and a uniform Ni–O bond length of (~1.97 Å).[47,52,54–56] However, first-principles calculations predict other space groups with lower symmetry to be more stable.[24,25,41,57] Sicolo et al. concluded that the experimentally observed symmetry is due to thermal fluctuations in which JT-distortions reorient continuously in the $P2_1/c$ space group.[25] Our results for the relative stability of $LiNiO_2$ are in agreement with Chen et al. for PBE+U.[41] The energies of the different space groups are within the thermal energy at room temperature (~25 meV/f. u for PBE+U with U = 5 and 6 eV) so that they can potentially be accessed at room temperature. Our results for the relative stabilities of the space groups are in agreement with those by Sicolo et al. when we utilize PBE+U with U = 5 eV.

The stable phases and Li orderings that we found by combining all delithiation products from the three space groups (**Figure 3**) are in agreement with those found by Mock et al.[48] using X-ray diffraction and Monte Carlo simulations, except for a phase at $x$ = 0.4 that Mock et al. reported to be stable while it is not in our calculations. This discrepancy might be due to the Hubbard-U correction that we employed, since the phase is stable without a +U correction, in agreement with previous computational results by Arroyo y de Dompablo et al.[58]

Even though the $LiNiO_2$ structure with the $P2_1/c$ space group (zig-zag ordering of JT-distorted $Ni^{3+}$ sites) is among the plausible ground-state candidates, none of its delithiation products is predicted to be a 0K-ground-state stable phase (**Figure 3**). I.e., none of the phases stable at 0 K exhibits the characteristic zig-zag JT ordering. A reason might be that, as the $P2_1/c$ space group is delithiated, the oxidation of $Ni^{3+}$ atoms to $Ni^{4+}$ hinders the thermal reorientation of JT-distorted $Ni^{3+}$ sites because of the short and immobile Ni-O bonds of $Ni^{4+}$. Hence, $Ni^{4+}$ destabilizes the JT zig-zag pattern. Radin et al.[24] concluded that the most stable JT-orderings permit an oxygen coordination in which each oxygen atom is part of exactly two short Ni-O bonds and one long Ni-O bond, thereby avoiding under and over-coordinated. Following this argument, the formation of a $Ni^{4+}$ forces 6 Ni-O bonds to shrink (to ~1.89 Å), and as a consequence, each of those 6 oxygen atoms becomes part of one more long and one more short Ni-O bond.

The energetics of TM migration in layered LiTMO$_2$ strongly depend on the oxidation state of the TM that determines the *Ligand Field Stabilization Energy* (LFSE) and the size of the tetrahedral sites onto which the TM atoms migrate.[3,11,59] In Li$_{0.5}$NiO$_2$, the Ni oxidation state and the tetrahedral sites are not favorable for Ni migration.[60] Ni in tetrahedral sites (Ni$_{tet}$) is generally in a 2+ oxidation state; to our knowledge no other oxidation state for Ni$_{tet}$ has been reported, nor does it appear in our calculations. In the P2/m space group, a Ni$_{tet}^{2+}$ oxidation state can only be achieved via charge disproportionation involving the migrating Ni atom and one of its neighbors (2 Ni$_{oct}^{3+}$ → Ni$_{tet}^{2+}$ + Ni$_{oct}^{4+}$). This oxidation process destabilizes the charge ordering and requires electron transfer from an e$_g$ level to a t$_{2g}$ level, going from a JT-distorted Ni$^{3+}$ to a Ni$^{4+}$ (2t$_{2g}^6$e$_g^1$ → t$_{2g}^6$ + e$^4$t$_2^4$). The change in LFSE in Li$_{0.5}$NiO$_2$ was calculated to be 6/5Δ$_o$ − 4/5Δ$_t$ +α whereas in Li$_{0.5}$MnO$_2$ it is equal to α, so that the energy of the charge transfer process is much higher in Li$_{0.5}$NiO$_2$.[10] Here, α is the energy splitting of the e$_g$ levels due to the JT distortion. **Figure S6** shows the formation energy for the Li-Ni dumbbell configuration for different Ni oxidation states together with the change in the molecular orbital diagram and the LFSE.

The above analysis of the electronic configuration corroborates our result that the formation of the partially inverse spinel or large domains of this phase in Li$_{0.5}$NiO$_2$ is energetically unlikely as its formation energy is high. There is no thermodynamic driving force, and the partially inverse spinel is rather unstable. In comparison, in Li$_{0.5}$MnO$_2$, the partially inverse spinel forms readily at room temperature in an energetically downhill process.[12,31] To our knowledge, the formation of the partially-inverse spinel has not been experimentally observed for Li$_{0.5}$NiO$_2$. Zhang et al. argued that the partially inverse spinel in LiNi$_{0.8}$Co$_{0.15}$Al$_{0.05}$O$_2$ (NCA) can't be captured because of its short life span.[17] Our calculations for Li$_{0.5}$NiO$_2$ suggest instead that the formation of the partially-inverse spinel cannot occur for Ni-rich LiTMO$_2$ because of unfavorable energetics. Ni in tetrahedral sites are unstable and would rapidly reverse to their original site or migrate further to their octahedral spinel site (forming spinel defects) before a significant fraction of other Ni atoms migrate to tetrahedral sites. Hence, the partially-inverted spinel will never be visited upon heating of Li$_{0.5}$NiO$_2$. In addition, NCA contains Co and Al, two dopants that have been argued to raise the energy needed for Ni migration,[8,9,61,62] especially the redox-inactive Al$^{3+}$ cannot donate electrons to neighboring Ni$^{3+}$ and thus hinders the required disproportionation reaction described above. From an electronic structure point of view, forming the partially inverse spinel phase requires net energy for raising electrons into higher electronic levels.

At a Li composition of $x = 0.5$, the inverse spinel (Ni)[LiNi]O$_4$ has previously been computationally investigated.[7] Here, we did not consider this phase given that it also has never been experimentally observed, and its formation energy is too high to be relevant for LiNiO$_2$ cathode materials.

Our results for the layered-to-spinel transformation are in agreement with Dahn et al.[18] and Goodenough et al.,[18,19] who previously reported that this phase transition requires high thermal energies. In addition to the energetics of the transformation, we previously determined the first-principles energy barrier to be at least 1.5 eV/defect.[63] For the transformation to occur, large domains of the crystal must concertedly undergo the transformation, which is statistically unlikely and further increases the observed phase-transition temperature.

Our results show that the energy barrier of the layered-to-spinel transformation depends on the space group of the layered structure. The differences in energy barrier arise from the entangled electron transfer mechanisms that must occur in the P2/m → Imma and the P2/c → P4$_3$32 transformations. In the P2/m → Imma transformation, all Ni atoms undergo the same electron transfer mechanism, whereas in the P2 → P4$_3$32 transformation, the charge transfer mechanism depends on the oxidation state of the migrating Ni atom in the layered structure.

While we have not investigated the layered-to-spinel transformation at concentrations other than $x = 0.5$ in Li$_x$NiO$_2$, we expect it to be most facile at the ideal spinel composition. The formation of the spinel structure for $x > 0.5$ might not be observed because of the high electrostatic repulsion of the additional Li atoms. In the case of $x < 0.5$, the transformation will likely be limited by a lack of mobile Ni$^{2+}$ and tetrahedral Li atoms to stabilize the structure.[64,65]

In the present work, we focused exclusively on the layered-to-spinel formation in the bulk of LiNiO$_2$. Even at room temperature spinel formation can occur in the surface regions of LiNiO$_2$ particles,[66–68] where the material may interact with its surroundings, including the electrolyte.[69] For example, oxygen vacancies in the surface have been linked with spinel formation, as they facilitate additional Ni-migration mechanisms.[66,70]

At zero Kelvin, our calculations predict a charge disproportionated structure (P2) for layered Li$_{0.5}$NiO$_2$ containing Ni$^{2+}$ and Ni$^{4+}$ but no Ni$^{3+}$ to be more stable than one containing JT-distorted Ni$^{3+}$ sites (P2/m), even though experimental data agrees with the P2/m structure.[48,51,71] Similarly, DFT predicts the charge-disproportionated P4$_3$32 symmetry for the spinel structure to be more stable than the JT-distorted Imma phase at zero Kelvin, but room-temperature experimental observations show the higher-symmetry Fd$\bar{3}$m phase group of the ideal spinel structure and no evidence for Ni$^{2+}$ in spectroscopy.[19,20,72,73] These discrepancies are reminiscent of the situation for fully lithiated LiNiO$_2$ and its apparent R$\bar{3}$m symmetry seen in diffraction experiments, and it could indicate that thermal effects at room temperature prevent charge disproportionation layered Li$_{0.5}$NiO$_2$ and spinel Li(NiO$_2$)$_2$. The room-temperature structures could be a thermal average of equivalent structures that differ in the orientation of the JT distortions. The ideal spinel structure with Fd$\bar{3}$m symmetry exhibits uniform Ni-O bond lengths and an average Ni oxidation state of (+3.5), and our DFT calculations predict robustly (i.e., independently of the U value) that such a structure is unstable. Kuwabara et al. pointed out that the heat of formation predicted by DFT is in better agreement with experimental measurements [71,74] when a charge-ordered spinel structure is taken as the computational model.[30] This further corroborates that the experimental observation is a thermal average.

## 5. Conclusion

We investigate the layered-to-spinel transformation in Li$_{0.5}$NiO$_2$ using first-principles calculations. We determined the likely structure of layered Li$_{0.5}$NiO$_2$ via a computational delithiation of fully lithiated LiNiO$_2$ with different symmetries, leading also to spinel structures with different symmetry space groups. Our calculations show that the phase transformation is a concerted process that requires nearly 20% of the Ni in the material to migrate simultaneously. Unlike Li$_{0.5}$MnO$_2$, the layered-to-spinel transformation of Li$_{0.5}$NiO$_2$ does not proceed via an intermediate partially-inverted spinel phase due to electronic frustration and an ordering of Jahn–Teller distortions. The space groups of the layered and spinel structure models have a significant impact on the energetics of the phase transition, underlining the importance of charge and Li/vacancy orderings for the properties of Li-Ni oxide phases. A comparison with spectroscopy results from the literature indicates that both the layered Li$_{0.5}$NiO$_2$ and the spinel Li(NiO$_2$)$_2$ phases observed in experiment are likely thermal averages of lower-symmetry structures with ordered Jahn–Teller distortions.

**Acknowledgments:** This work was supported by the Alfred P. Sloan Foundation Grant No. G-2020-12650. We acknowledge computing resources from Columbia University's Shared Research Computing Facility project, which is supported by NIH Research Facility Improvement Grant No.

1G20RR030893-01, and associated funds from the New York State Empire State Development, Division of Science Technology and Innovation (NYSTAR) Contract No. C090171, both awarded April 15, 2010. We thank John. H. Harding for providing $LiNiO_2$ structures.

# SUPPORTING INFORMATION

# Layered-to-Spinel Phase Transformation in $Li_{0.5}NiO_2$ from First Principles

C. Komurcuoglu[1], A. C. West[1] and A. Urban[1]

[1]Department of Chemical Engineering and Columbia Electrochemical Energy Center, Columbia University

E-mail: au2229@columbia.edu

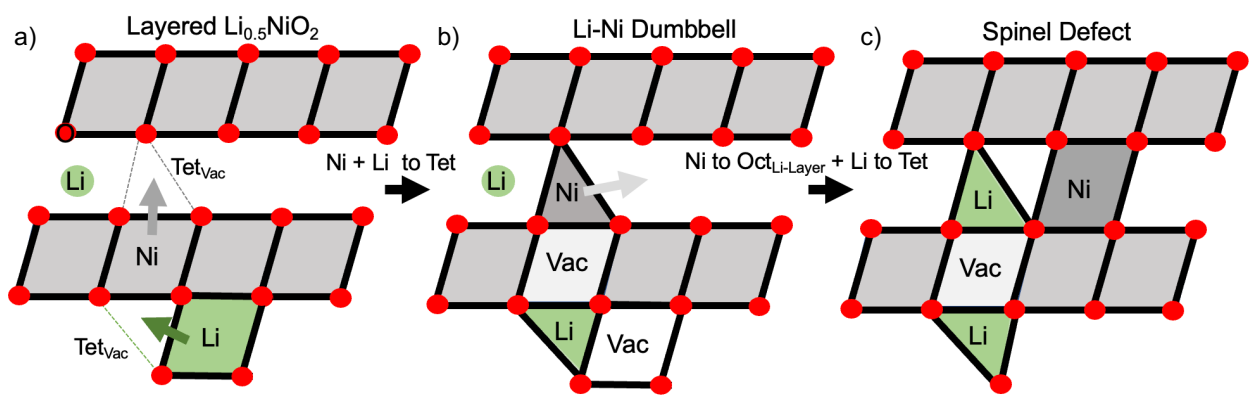

**Figure S1**: Schematic of the Ni migration mechanism in $Li_{0.5}NiO_2$. **a)** Layered structure containing vacancies on which Li and Ni migrate. **b)** The Li-Ni dumbbell, the building block of the partially-inverse-spinel structure. **c)** The spinel defect, the building block of the spinel structure composed of $Ni_{oct,\ Li-layer}$ + $Li_{tet,\ Li-Layer\ 1}$ + $Li_{tet,\ Li-Layer\ 2}$.

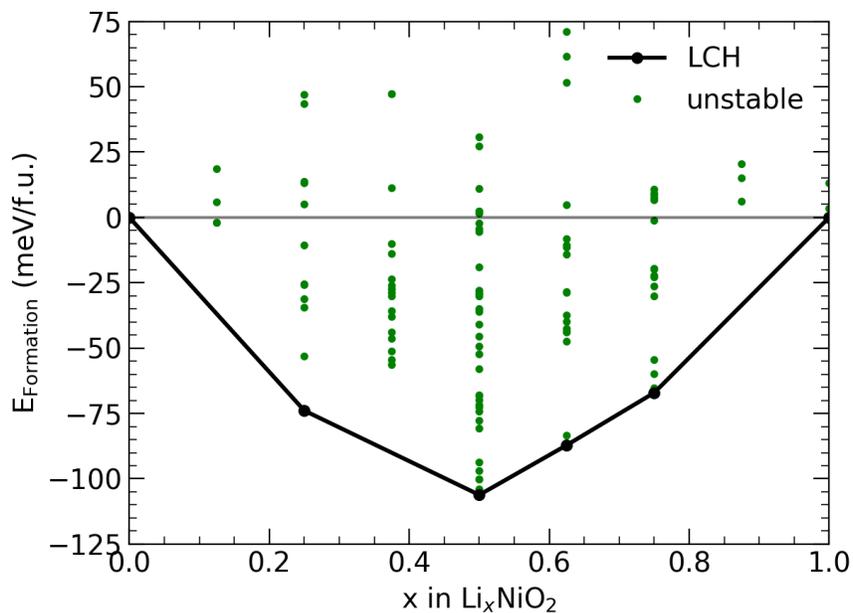

**Figure S2:** $Li_xNiO_2$ Lower Convex Hull, $E_{Formation}$ vs. Li content, showing all comuted stable and unstable configurations.

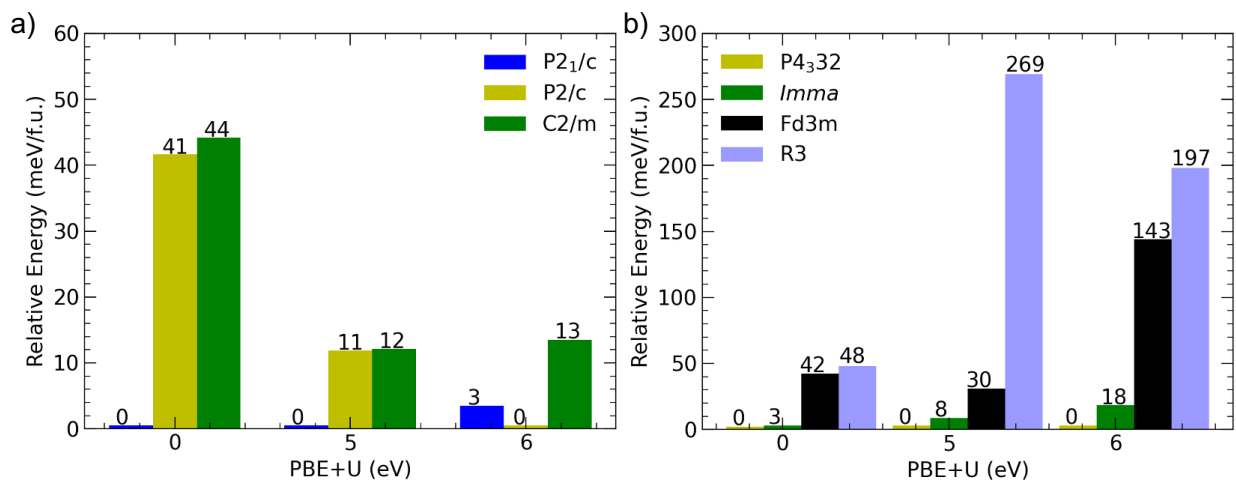

**Figure S3:** Relative formation energies of layered $LiNiO_2$ and spinel $Li(NiO_2)_2$ for different U-values and space groups. **a)** Relative stabilities of layered $LiNiO_2$ structures with different space groups as a function of the U value. **b)** Formation energies of spinel structures with different space groups relative to the energy of the $P4_332$ space group, which is the most stable for any choice of U value.

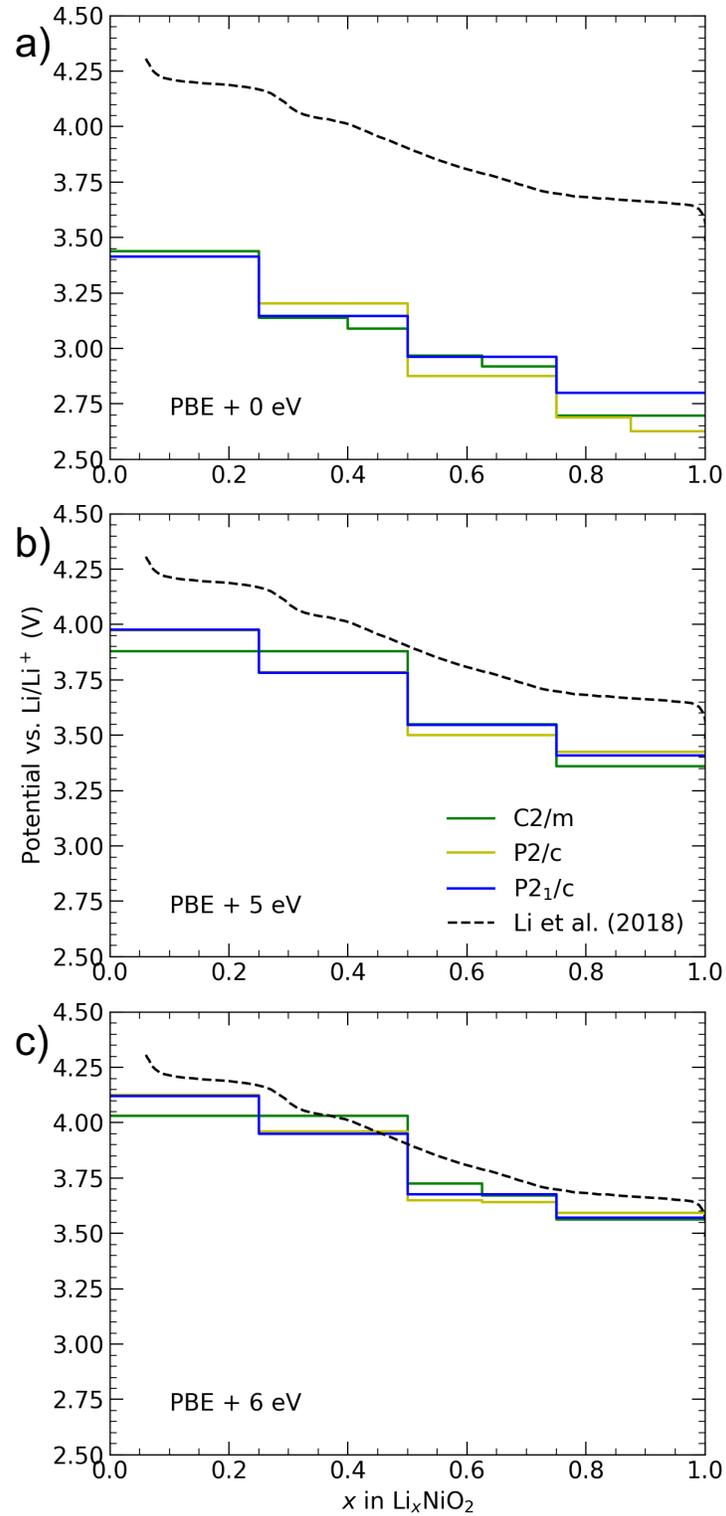

**Figure S4**: Voltage profiles for the delithiation of LiNiO$_2$ with 3 different initial space groups ($C2m$, $P2/c$, and $P2_1/c$) and 3 different choices of U values. Experimental V-Profile replotted from Li et al.[47]

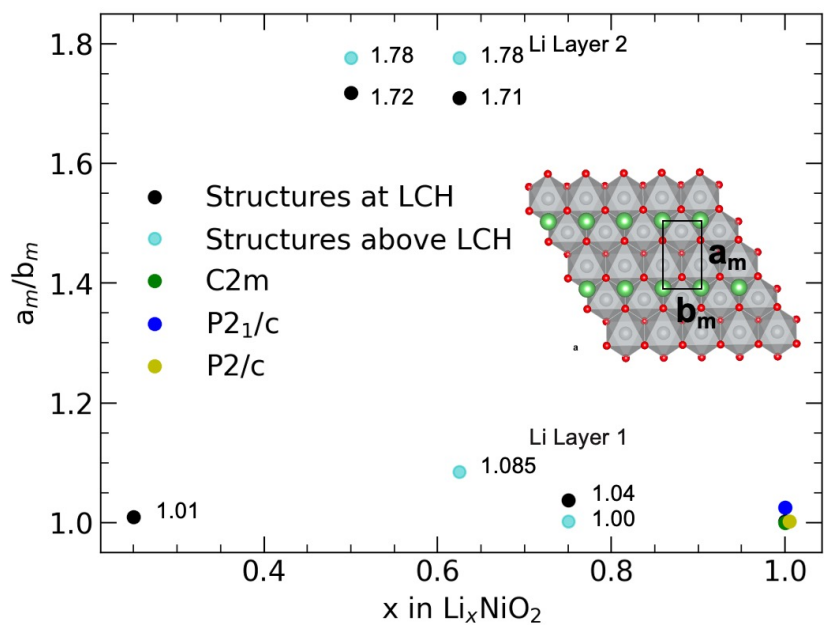

**Figure S5:** Calculated a/b ratios for ordered stable phases and structures slightly above the hull from **Figure 4** in the main manuscript. For $x = 1.0$ the values are shown for 3 space groups of $LiNiO_2$.

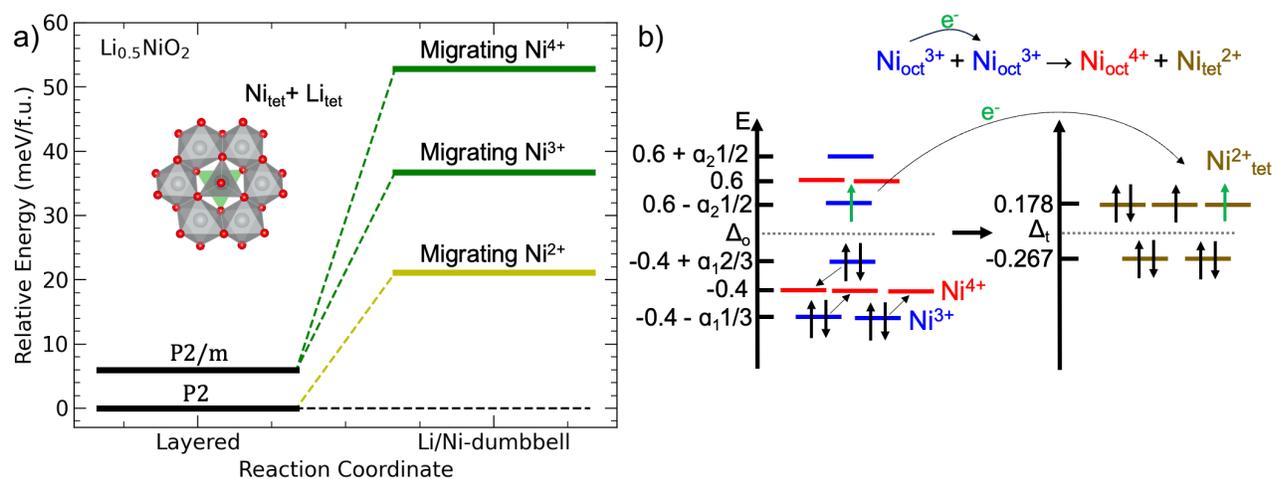

**Figure S6: a)** Li-Ni dumbbell formation energy for different migrating Ni oxidation states and **b)** schematic of the electronic structure change upon $Ni^{3+}$ migration. One octahedral $Ni^{3+}$ atom raises the $t_{2g}$ electrons into tetrahedral $e_g$ and $t_{2g}$ levels as it oxidizes a $Ni^{3+}$ neighbor. $Ni^{4+}$ must oxidize two $Ni^{3+}$ to be in the required 2+ oxidation state. $Ni^{2+}$ migration does not require any electron transfer.

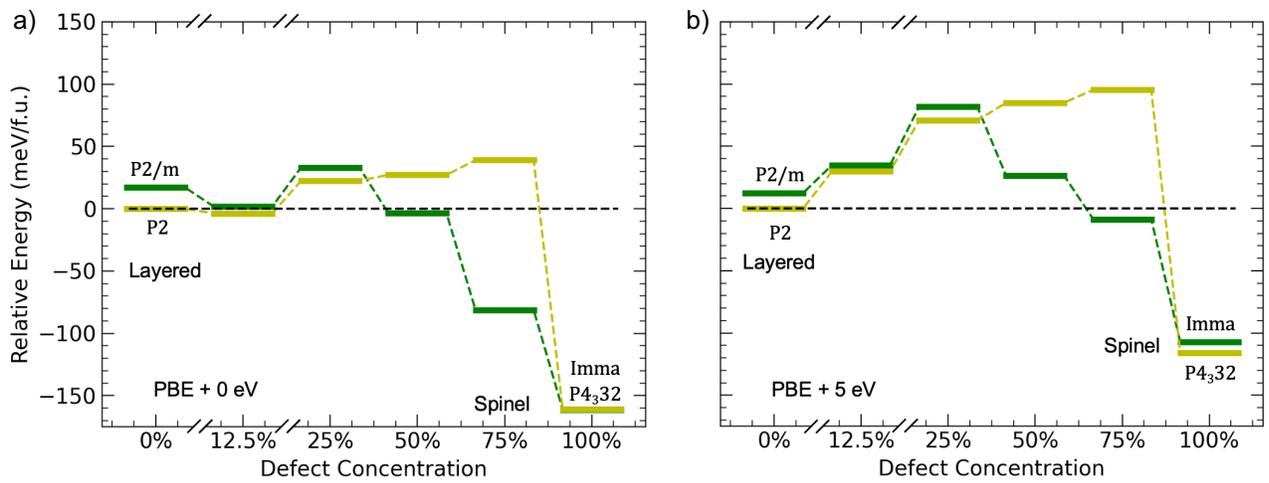

**Figure S7**: Energetics of the layered-to-spinel transformation as a function of the spinel defect concentration for two layered $Li_{0.5}NiO_2$ (space groups $P2/m$ and $P2$) and for two U values. **a)** PBE (U = 0 eV) and **b)** PBE+U (U = 5 eV).